\documentclass[11pt,a4paper]{article}

\usepackage{a4wide}
\usepackage{amsmath}
\usepackage{amssymb}
\usepackage{booktabs}
\usepackage{graphicx}
\usepackage{hyperref}
\usepackage{xcolor}
\usepackage[utf8]{inputenc}

\newcommand{\tbl}[1]{\caption{#1}\vspace{0.5em}\centering}
\newcommand{\colrule}{\midrule}
\newcommand{\botrule}{\bottomrule}
\newcommand{\chapter}[2][]{}
\newcommand{\comment}[1]{}
\begin{document}

\title{\textbf{The strong coupling: a theoretical perspective}}
\author{Gavin P.\ Salam\footnote{On leave from CNRS, UMR 7589, LPTHE, F-75005, Paris, France.}\\
  Theoretical Physics Department, CERN, CH-1211 Geneva 23, Switzerland
}
\date{}
\maketitle

\vspace{-12em}
\begin{flushright}
  CERN-TH-2017-268
\end{flushright}
\vspace{7.5em}

\begin{abstract}
  This contribution to the volume ``\emph{From My Vast Repertoire ---
  The Legacy of Guido Altarelli}'' discusses the state of our
  knowledge of the strong coupling. 
\end{abstract}
\vspace{0.5em}

\section{Introduction}
% ----------------------------------------------------------------------
\chapter[The strong coupling: a theoretical perspective]{The strong
  coupling:\\ a theoretical perspective}
\newcommand*{\as}{\alpha_s}
\newcommand*{\mz}{m_Z}
\newcommand*{\asmz}{\as(\mz)}
\renewcommand*{\comment}[1]{\textbf{\color{red}[#1]}}
\newcommand{\order}[1]{{\cal O}\!\left(#1\right)}
\newcommand*{\GeV}{\,\text{GeV}}
\newcommand*{\TeV}{\,\text{TeV}}
\newcommand*{\ycut}{y_\text{cut}}
% command to help insert elements just for the arXiv version
\newcommand*{\arxivonly}[1]{%
  \ifcsname arxivversion\endcsname%
  #1
  \else%
  \fi%
}

The strong coupling, $\as$, is one of the fundamental parameters of
the Standard Model.
It enters into all cross section calculations for processes at the
Large Hadron Collider (LHC), whether directly at leading order, or
through higher-order QCD calculations.
It also enters indirectly through the evolution of parton distribution
functions (PDFs) and their correlation with the strong coupling.
Consequently, as the LHC experiments' work evolves towards precision
physics, accurate knowledge of the strong coupling is becoming
increasingly important.
The value of the coupling matters also for the question of gauge
coupling unification at high scales and for the stability of the
universe in any given particle-physics scenario.

The question of the value of $\as$ is hotly debated, with a range of
discussions in the
literature~\cite{Bethke:2011tr,Pich:2013sqa,Moch:2014tta,dEnterria:2015kmd,Patrignani:2016xqp}.
It's a subject that Guido had an active interest
in~\cite{Altarelli:2013bpa}.
Since, with Siggi Bethke and G\"unther Dissertori, I'm one of the authors
of the PDG review chapter on QCD, which also includes a discussion and
average of $\as$, we often had exchanges on the subject.
As I'll explain below, Guido was rather critical of our PDG approach
to the question.

To illustrate the problem of establishing the value of $\as$, consider
the two following determinations: one, from a lattice-QCD calculation
of Wilson loops, quotes
$\asmz = 0.11840 \pm 0.00060$~\cite{McNeile:2010ji};
another, from a fit to the thrust distribution in $e^+e^-$ 
collisions~\cite{Abbate:2010xh}, yields $0.1135\pm0.00105$.
The two determinations are four standard deviations apart from each
other,
and there is no single value of $\as$ that isn't at least three
standard deviations from one or other of them.
Both determinations pay extensive attention to the question of
potential systematic uncertainties, yet
in all likelihood, at least one of the two has underestimated them.

A first question is what accuracy do we need for the strong coupling.
Consider the case of LHC phenomenology.
The change induced in key LHC cross sections, e.g.\ the top-quark
cross section or the Higgs-boson cross section, in going from
$\as=0.118$ to $\as=0.113$ is a reduction of about
$8{-}9\%$.\footnote{This is based on the CT14nnlo PDF
  set~\cite{Dulat:2015mca}, for a centre-of-mass energy of $\sqrt{s} =
  13\TeV$, taking into account the correlation of the
  PDFs with $\as$.
  Cross sections were evaluated with the
  ggHiggs code~\cite{Bonvini:2016frm} (at N3LO~\cite{Anastasiou:2016cez}) and
  the top++ code~\cite{Czakon:2011xx} (at NNLO).
  In the case of $t\bar t$ production with ABMP16
  PDFs~\cite{Alekhin:2017kpj}, the effect is larger, a $14\%$
  reduction.
  %
  % ------------ 13 TeV Higgs ---------------------
  % data/higgs-xsc/ct14nnlo_as_0118.dat:NNNLO = 46.4074
  % data/higgs-xsc/ct14nnlo_as_0113.dat:NNNLO = 42.0545
  % Ratio = 0.906
  %
  % ABMP16:
  % data/higgs-xsc/ABMP16als118_5_nnlo.dat:NNNLO = 46.6645
  % data/higgs-xsc/ABMP16als113_5_nnlo.dat:NNNLO = 42.5446
  %------------ 13 TeV CT14 ttbar ---------------------
  %
  % ttbar/sigma-results/LHC13000-NNLO-CT14nnlo-PDFErr.res
  % # mt   central   pdf_min   pdf_max
  % 173.2  807.296  771.743  842.277
  % sigma-results/LHC13000-NNLO-CT14nnlo_as_0113.res
  % # mt   central   scale_min   scale_max
  % 173.2  741.043  741.043  741.043
  %
  % RATIO = 0.9179
  %
  %------------ 8 TeV CT14 ---------------------
  % ttbar/sigma-results/LHC8000-NNLO-CT14nnlo-PDFErr.res (as=0.118)
  %  # mt   central   pdf_min   pdf_max
  %   173.2  246.652  232.538  262.821  
  % sigma-results/LHC8000-NNLO-CT14nnlo_as_0113.res 
  % # mt   central   scale_min   scale_max
  % 173.2  224.98  224.98  224.98
  %
  % RATIO: .85496
  %
  %------------ 8 TeV ABMP16 ---------------------
  % ttbar/extra-sigma/LHC8000-NNLO-ABMP16als118_5_nnlo-PDFErr.res
  % # mt   pdf member   muF   muR   alpha_s   sigma_tot
  % 173.2        0        1        1        0.107581        234.936
  %
  % ttbar/extra-sigma/LHC8000-NNLO-ABMP16als113_5_nnlo-PDFErr.res
  % # mt   pdf member   muF   muR   alpha_s   sigma_tot
  % 173.2        0        1        1        0.103439        200.99
  %
  % RATIO: .85496
}
This is larger than the total theoretical uncertainties on these cross
sections from missing higher-order corrections (about
$4-6\%$~\cite{deFlorian:2016spz,Anastasiou:2016cez}) and larger also
than current or foreseen experimental uncertainties: the top cross
section is measured to about $3-4\%$
uncertainty~\cite{Aad:2014kva,Khachatryan:2016mqs}, and in the long
term the Higgs cross section should also reach a similar or better
precision.
Even a $1\%$ uncertainty on $\as$ leads to effects that are comparable
to any other single theoretical uncertainty on the Higgs cross
section, i.e.\ at the $2\%$ level.
Therefore, only for a determination of the coupling with a precision
comfortably below the percent level can $\as$ uncertainties largely be
ignored for extracting fundamental information from the
LHC.\footnote{One may also consider how $\as$ impacts vacuum stability
  estimates, assuming validity of the Standard Model up to high
  scales.
  A $1\%$ increase in $\asmz$ has roughly the same impact on the
  stability criterion as a $0.4\GeV$ decrease in the top-quark mass,
  while a $5\%$ increase in the value of $\asmz$ would ensure
  stability, rather than just metastability, of the Standard Model
  vacuum~\cite{Espinosa:2015qea}.}

To understand the limitations that arise in determining the strong
coupling, it is useful to keep in mind the essence of any determination
of the coupling.
Some quantity $V$ is measured experimentally, giving a result
$V_\text{exp} \pm \delta V_\text{exp}$.
This needs to be related to a theoretical prediction for the same
quantity in terms of powers of the coupling,
\begin{equation}
  \label{eq:1}
  V_\text{th}(\as(\mu), \mu) = \sum_n^N c_n(\mu) \as^n(\mu)
  + \order{\as^{N+1}}
  + \order{\frac{\Lambda^p}{Q^p}}\,,
\end{equation}
where the $c_n$ factors are the coefficients of the perturbative
series, which can in practice be calculated up to some finite order
$n=N$.
They depend on the choice of renormalisation scale $\mu$, as does the
coupling itself.
The quantity $\Lambda$ is the non-perturbative scale of QCD and $Q$ is
the order of magnitude of the momentum transfer in the process used
for measuring $V$.
The term $\frac{\Lambda^p}{Q^p}$ reflects the inevitable existence of
non-perturbative contributions.
Our understanding of its structure and relevance is, in some contexts,
the subject of debate, though in most cases at the least the value of
the power $p$ is known.

Requiring $V_\text{th}(\as(\mu),\mu) = V_\text{exp}$ in Eq.~(\ref{eq:1}) fixes
$\as$.
The uncertainty on the $\as(\mz)$ determination then has three sources:
(i) the extent to which $V$ can be measured precisely, i.e.\ the size
of the $\delta V_\text{exp}$ uncertainty;
(ii) the estimated impact of terms $c_n$ beyond those that can be
calculated with today's technology, of order $\as^{N+1}$, e.g.\ as
found by varying $\mu$;
and (iii) the size of the ``power correction'' terms, $\Lambda^p/Q^p$ and
the degree to which they can be reliably understood.
There may also be missing higher-order electroweak terms, or
uncertainties associated with other fundamental parameters.
The discussion of different determinations will essentially be a
discussion of the relative sizes of each of these sources of
uncertainty, and the degree of consensus on our understanding of them.

The Particle Data Group (PDG)~\cite{Patrignani:2016xqp} world average
(in the QCD review chapter) limits its inputs to cases where the
perturbative series is known at least to next-to-next-to-leading order (NNLO),
and in the interests of brevity the discussion below will similarly
concentrate on those cases.
In the PDG we have taken the approach that we should be as neutral as
possible with regard to disputes in the community about different
determinations, with uniform prescriptions applied to all reasonable
determinations.
That is motivated in part by a desire to minimise any risk of bias in
the outcome of the average.
I think it's fair to say that Guido wasn't impressed by this approach.
In our discussions on the subject he would insist that one should
attempt to bring a theorist's critical view to each determination.
He argued
\begin{quotation}
  \noindent [...] one should select few theoretically simplest processes
  for measuring $\as$ and consider all other ways as tests of the theory.
\end{quotation}
In the chapter I'll give my take on what is needed to make a ``clean''
determination of the strong coupling: clean (and ``transparent'') were
words regularly used by Guido in this context.

%----------------------------------------------------------------------
\section{Jet rates and event shapes and  in $e^+e^-$ collisions}

One natural way of thinking about the meaning of the QCD coupling is
that it governs the probability of emitting a gluon.
Gluons, of course, cannot be directly observed, but jets of hadrons
can be used as a stand-in for hard gluons, i.e.\ for gluons that are
energetic and radiated at large angles with respect to their emitter.

The simplest environment in which to study jets is $e^+e^- \to
\text{hadrons}$ reactions.
Perturbatively, the lowest order process is $e^+e^- \to q\bar q$,
i.e.\ a two-jet event.
There is a probability $\as$ to radiate a hard gluon, giving
$e^+e^- \to q\bar q g$, i.e.\ a three-jet event.
By measuring the fraction of three-jet events one can determine $\as$.

There is considerable freedom in how one carries out the extraction:
there are different algorithms for defining the jets, including the
Durham~\cite{Catani:1991hj} and Cambridge~\cite{Dokshitzer:1997in}
algorithms.
Each algorithm comes with a parameter to define how energetic the
emission should be in order to be considered a jet.
For the Durham and Cambridge algorithms, the parameter is called
$y_\text{cut}$, and corresponds to the squared transverse momentum of
the emission, normalised to the squared centre-of-mass energy.

\begin{figure}[tp]
  \centering
  \includegraphics[width=0.49\textwidth,page=9]{figs-salam/y3C-OPAL-v-NNLO.pdf}%
  \hfill%
  \includegraphics[width=0.49\textwidth]{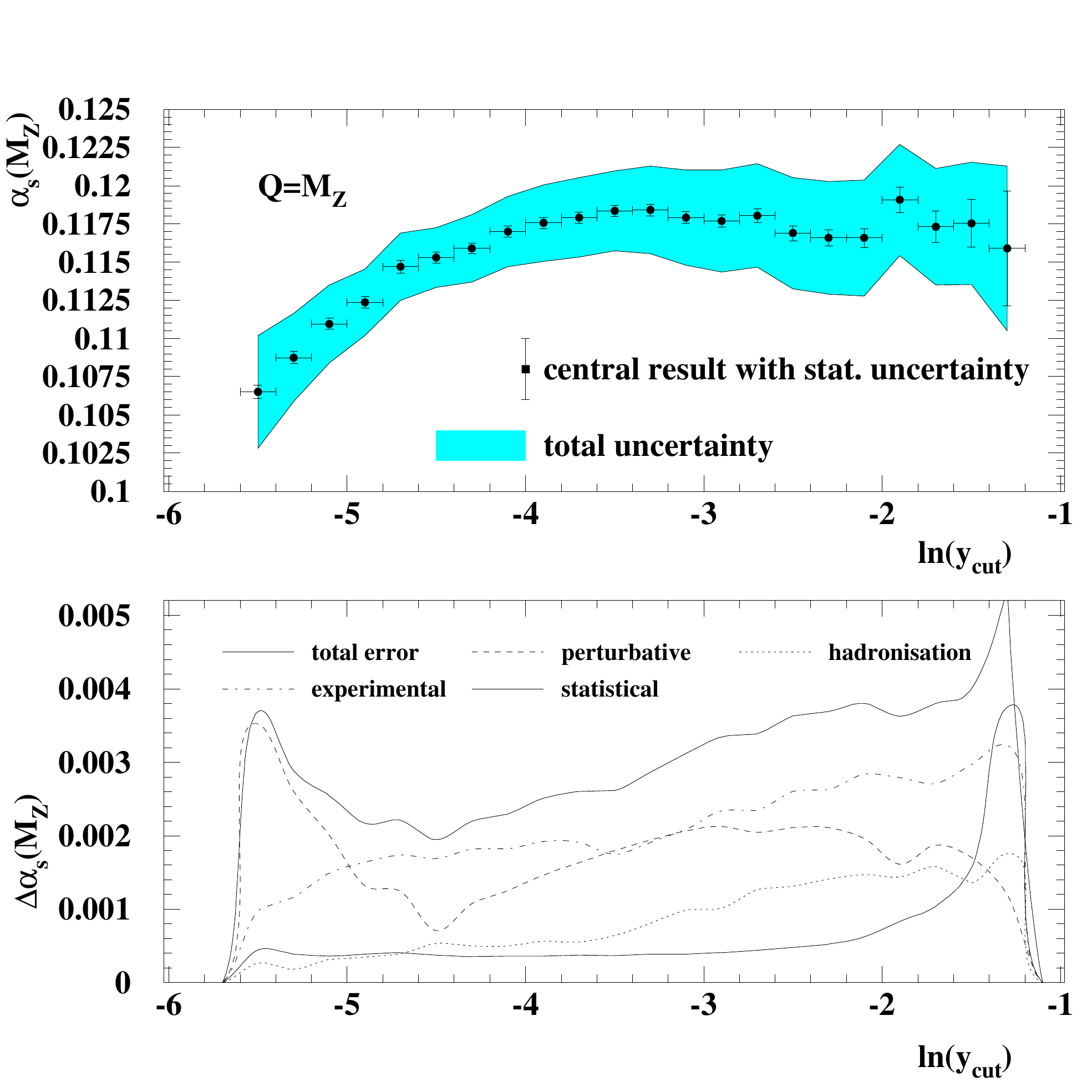}%
  \\
  \includegraphics[width=0.49\textwidth,page=10]{figs-salam/y3C-OPAL-v-NNLO.pdf}%
  \hfill%
  \includegraphics[width=0.49\textwidth,page=4]{figs-salam/y3C-OPAL-v-NNLO.pdf}%
  \caption{Top left: comparison of ALEPH jet rate
    data~\cite{Heister:2003aj} for the Durham algorithm
    with the pure NNLO result for
    $\as(\mz)=0.118$~\cite{Weinzierl:2010cw} (see also
    \cite{GehrmannDeRidder:2008ug,DelDuca:2016csb}), and the NNLO
    result multiplied by 
    hadronisation corrections estimated with
    Pythia~6.428~\cite{Sjostrand:2006za} with the DW tune.
    Top right: strong coupling (top) and the uncertainty breakdown from
    the ALEPH jet rate extraction (Fig.~1 of
    Ref.~\cite{Dissertori:2009qa}).
    Bottom left: ratio of NNLO theory to data for $\asmz=0.118$, with
    the same data and simulation choices as for the top-left
    plot.
    Bottom right: ratio of NNLO theory to data for $\asmz=0.118$, but
    now with the Cambridge jet algorithm, OPAL data and
    Pythia~8.223~\cite{Sjostrand:2014zea} (Monash 2013
    tune~\cite{Skands:2014pea}) for the hadronisation.  }
  \label{fig:aleph-jet-rate}
\end{figure}

To illustrate some of the characteristics and challenges of
extractions of the coupling from jet rates,
Fig.~\ref{fig:aleph-jet-rate} (top-left) shows the Durham 3-jet rate
as a function of $y_\text{cut}$, 
as measured by the ALEPH collaboration at particle (i.e.\ hadron)
level.
It is compared to the pure NNLO prediction in 5-flavour massless QCD
at parton level and to the NNLO result multiplied by an additional
hadronisation correction.
The upper axis shows the effective transverse momentum ($p_t$) cut
that a given value of $\ln y_\text{cut}$ corresponds to.
For a typical choice $\ln y_\text{cut} = -4.0$, one hovers close to a
$p_t$ cut of $10\GeV$.
If one considers that a jet's energy may change by an amount of the
order of a GeV due to the parton to hadron transition, then the fact
that the jet $p_t$ is just $10\GeV$ becomes a concern.
That parton-to-hadron transition is believed to be the main reason the
NNLO (parton) theory doesn't immediately agree with the hadron-level
data.
To extract a value of the coupling it is mandatory to apply a
hadronisation correction to the NNLO prediction (and also corrections
for $b$-quark mass effects).
This was done for the analysis in the top right-hand plot of
Fig.~\ref{fig:aleph-jet-rate}, using Monte Carlo event generators to
estimate the hadronisation correction (about a $5\%$ effect).
The plot shows the ALEPH collaboration's resulting extraction of the
strong coupling as a function of
$y_\text{cut}$~\cite{Dissertori:2009qa}.
For $\ln y_\text{cut} > -4$ the extracted $\as$ value is fairly
independent of $\ycut$, a sign of robustness of the analysis.
Below that, however, the result depends substantially on the choice of
$y_\text{cut}$.
One can attribute that feature to a breakdown of fixed-order
perturbation theory, associated with logarithmically enhanced terms of
the perturbative series, which go as $(\as \ln^2 y_\text{cut})^n$.
%%-- Other thinking
% OPAL + JADE 3-jet rate 
% https://hepdata.net/record/ins513337 
% Table 43, Cam alg, 91 GeV
% (see also data/JADEOPAL_2000_I513337-d43-x01-y02.yoda)
%
% ycut = 0.0422: 3-jet rate(%) = 12.11 ± 0.05 (stat) ± 0.72 (sys)
% ycut = 0.0562: 3-jet rate(%) =  9.37 ± 0.05 (stat) ± 0.55 (sys)
% 
% i.e. 0.6% syst error and stat error that's negligible
%
% What kt does this correspond to?
%       ycut = 0.0422 -> 18.73 GeV  [ = sqrt(0.0422) * 91.2 ]
%       ycut = 0.0562 -> 21.62 GeV
% 
% A 1 GeV shift is 5.3% of the pt
% A 3 GeV shift is 30% rate change
% So 1 GeV uncertainty is 10% rate change = 10% of alpha_s...
%
% JADE paper, 1205.3714 shows hadron/parton < 1 (corresponding to jet
% losing energy) -- ratio is about 0.87 at 43.8 GeV, cf. Fig.8

The final extraction from the ALEPH collaboration involves a choice of
$\ln y_\text{cut} \simeq -3.9$ that remains within the plateau and
minimises the final error.
It gives:
\begin{align*}
  \asmz =
  0.1175 &\pm 0.0004 \text{(stat.)}
         \pm 0.0016 \text{(det.)}
         \pm 0.0011 \text{(exp.)}\nonumber\\
         &\pm 0.0006 \text{(had.)}
         \pm 0.0002 \text{(mass)}
         \pm 0.0014 \text{(pert.)}\,,\\
  %0.1175 \pm 0.0020 \text{(exp)} \pm 0.0015 \text{(theo)}
  =0.1175 &\pm 0.0025 \text{(total)}\,,
\end{align*}
for which the dominant quoted error sources are detector and other
experimental systematics as well as missing high-order contributions
(assessed through variation of the renormalisation scale).

The aspect of this kind of determination that is perhaps most called
into question is the hadronisation correction, because the way
non-perturbative effects arise in Monte Carlo simulations (through a
cutoff) does not match the way they need to be applied to perturbative
calculations.
Rather than applying a cutoff at some scale of order $1\GeV$,
perturbative calculations integrate down to zero momentum, but using
an integrable, perturbative expansion of the coupling.
The overall size of the hadronisation correction is $4-5\%$, as is
visible from the ratio plot in the bottom-left of
Fig.~\ref{fig:aleph-jet-rate}.
The definition of how hadronisation interfaces with a perturbative
calculation could conceivably modify this by a couple of percent.

A further question with jet-rate data is the stability of the
determination.
The bottom-right plot of Fig.~\ref{fig:aleph-jet-rate} is analogous to
the bottom-left one except that the OPAL data replaces the ALEPH data,
the jet algorithm has been switched to the Cambridge algorithm and
hadronisation corrections have been evaluated with the Pythia~8
generator instead of Pythia~6.
The ratio of theory to data goes up by over $6\%$, with each of the
three changes accounting for about one third of that.\footnote{
The OPAL collaboration had already noted the smaller value of $\as$
for the Cambridge algorithm than for the Durham algorithm in fits at
NLO+NLL~\cite{Abbiendi:2005gb}.}
Since the OPAL data have larger systematic uncertainties than the
ALEPH data, the results appear to be still (just barely) compatible
with $\asmz=0.118$.\footnote{Note that in the full OPAL fits of
  Ref.~\cite{Abbiendi:2005gb}, which use slightly different jet
  variables than the $3$-jet rate discussed here, the experimental
  systematic error on $\as$ is substantially smaller than one might
  deduce based on the size of the band in the bottom-right plot of
  Fig.~\ref{fig:aleph-jet-rate}.
}
Nevertheless the difference between the bottom left and right-hand
plots of Fig.~\ref{fig:aleph-jet-rate} could be taken as a more
conservative estimate of the possible size of theoretical
uncertainties in $\as$ determinations from $e^+e^-$ jet rates.

\begin{table}
  \tbl{Determinations of the strong coupling from jet rates and
    event shapes in $e^+e^-$ collisions.}
  {\footnotesize
    \begin{tabular}{@{}lll@{}}
      \toprule
      Determination & Data and procedure & Reference\\
      \colrule
$0.1175 \pm 0.0025$ & ALEPH 3-jet rate (NNLO+MChad) & \cite{Dissertori:2009qa}\\[1pt]
$0.1199 \pm 0.0059$ & JADE 3-jet rate (NNLO+NLL+MChad) & \cite{Schieck:2012mp}\\[6pt]
%\midrule
$0.1224 \pm 0.0039$ & ALEPH event shapes (NNLO+NLL+MChad) & \cite{Dissertori:2009ik}\\[1pt]
$0.1172 \pm 0.0051$ & JADE event shapes (NNLO+NLL+MChad) & \cite{Bethke:2008hf}\\[1pt]
$0.1189 \pm 0.0041$ & OPAL event shapes (NNLO+NLL+MChad) & \cite{OPAL:2011aa}\\[6pt]
%\midrule
$0.1164\;^{+0.0028}_{-0.0026}$  & Thrust (NNLO+NLL+anlhad) & \cite{Davison:2008vx}\\[1pt]
$0.1134\;^{+0.0031}_{-0.0025}$ & Thrust (NNLO+NNLL+anlhad) & \cite{Gehrmann:2012sc}\\[1pt]
$0.1135 \pm 0.0011$ & Thrust (SCET NNLO+N$^3$LL+anlhad) & \cite{Abbate:2010xh}\\[1pt]
$0.1123 \pm 0.0015$ & C-parameter (SCET NNLO+N$^3$LL+anlhad) & \cite{Hoang:2015hka}\\[1pt]
      \botrule
    \end{tabular}} 
  \label{tab:event-shapes}
\end{table}

The ALEPH result is reproduced in table~\ref{tab:event-shapes}
together with a range of other strong-coupling determinations (all at
least at NNLO) from hadronic final-state measurements at LEP and
earlier $e^+e^-$ colliders.
As well as jet rates, they also make use of ``event shapes'' such as the
thrust~\cite{Brandt:1964sa,Farhi:1977sg},
$C$-parameter~\cite{Ellis:1980wv} or jet
broadenings~\cite{Catani:1992jc}.
Each event shape provides a continuous measure of the extent to which
an event's energy flow departs from a pure back-to-back (i.e.\ leading
order $e^+e^- \to q\bar q$) structure.
One sees that the central values and uncertainties for $\as$ vary quite
substantially between determinations, even though they all relate to
the same underlying phenomenon, the probability of gluon emission from
a quark-antiquark system.
This is perhaps unsurprising given what we've seen in
Fig.~\ref{fig:aleph-jet-rate}.

The result with the highest quoted precision is that for the thrust
(SCET),
\begin{equation}
  \label{eq:thrust-result}
  \as(m_Z) = 0.1135 \pm 0.0002_\text{exp.} \pm 0.0005_\text{hadr.} \pm
  0.0009_\text{pert.} = 0.1135 \pm 0.0011 \,,
\end{equation}
closely followed by the C-parameter determination.
A first point to be aware of is that in the 2-jet limit the thrust
and $C$-parameter are highly correlated, both in the data and in the
structure of the theoretical calculations. 
The two extractions are very similar in approach: they supplement
fixed-order perturbation theory with the resummation of enhanced
logarithmic contributions, specifically accounting for terms ranging
from $\as^n \ln^{n+1}$ down to $\as^n \ln^{n-2}$, i.e.\ N$^3$LL
accuracy.
Furthermore they use an analytic estimate of hadronisation corrections
with a fitted free parameter, which to a first approximation
corresponds to a shift of the distribution by a shift of the thrust
(specifically $1-T$) or $C$-parameter by an amount proportional to a
``power correction'' $\Lambda/Q$ (see the reviews
\cite{Beneke:1998ui,Dasgupta:2003iq}).
An advantage of analytic hadronisation estimates is that they can address
the concern of matching Monte Carlo hadronisation corrections to
perturbative calculations.

Guido's comment about the thrust result was
\begin{quote}
  I think that this is a good example of an underestimated error which
  is obtained within a given machinery without considering the limits
  of the method itself.
\end{quote}
What might these methodological limitations be?
One is that the formalism of resummation holds only for
$C, 1-T \ll 1$, where every emission is so soft and collinear that one
can effectively neglect the kinematic cross-talk (e.g. energy-momentum
conservation) that arises when there are multiple emissions.
There is almost no quantification in the literature of the corrections
induced by such cross-talk, so they potentially represent a neglected systematic
error.

Another limitation is that the power correction that is used holds in
the 2-jet limit, i.e.\ $1-T\ll 1$.
However the fits extend into the full 3-jet region, e.g.\
$(6\GeV)/Q<1-T<0.33$ for the thrust, keeping in mind that
$1-T = \frac13$ is the largest value that can be obtained with 3
partons.
One way of viewing the problem is that the power correction in this
region should be governed by a different operator and simply taking
the 2-jet result is risky.
A different way of phrasing this is that if the average effect of
hadronisation is essentially a shift of $1-T$ by an amount
$\delta \sim \Lambda/Q$, then $\delta$ itself is likely to be a
non-trivial function of the parton-level value of $1-T$.
Only for small values of $1-T$ may one neglect that functional
dependence.

A sign that this might be causing problems comes from the dependence
of the thrust fit results on the choice of limits.
Fig.~17 of Ref.~\cite{Abbate:2010xh} shows that modestly restricting
the fit range to $(8\GeV)/Q<1-T<0.25$ increases the central fit value
to about $0.1151$, an increase of $0.0016$ relative to
Eq.~(\ref{eq:thrust-result}), which is larger than the overall quoted
error of $0.0011$.\footnote{A full interpretation would require an analysis of the
  correlation of the errors of the results with different choices of
  limits.
  Interestingly, the corresponding danger sign is not there for the
  $C$-parameter results.
  However the absence of an obvious danger sign in the numerical fit
  should not be taken as an indication of the absence of danger.
  In particular, with Gionata Luisoni and Pier Monni, we have started
  investigating the impact on $\asmz$ of different forms of
  $C$-dependence for the power correction.
  Our preliminary finding is that the effect is substantial,
  potentially of the order of $5\%$.}

Yet another concern relates to the treatment of experimental
systematic errors.
The ALEPH fit had total detector and experimental systematic errors of
$0.0019$, to be compared to those quoted in the SCET thrust fit of
$0.0002$.
Even if the latter puts together several experimental results, such a
large reduction is puzzling.

What do I conclude about event shapes?
Ultimately I have doubts about how realistic it is to extract $\asmz$
with $1\%$ precision from data whose underlying physical scale is
$10{-}20\GeV$, using observables that have $\sim 1\GeV/Q$ power
corrections.
Perhaps, today, the value of such fits should instead be seen in terms
of what they might teach us about the limits of our understanding of
hadronisation corrections and also of resummation.

%----------------------------------------------------------------------
\section{LEP electroweak fits}

Rather than looking for the signature of actual gluon emission,
electroweak (EW) fits for $\as$ rely on the slight non-cancellation
between higher-order real and loop graphs in a range of EW
observables, many of them connected with $Z$ production at LEP and
SLC.
For example the quantity
$R = \frac{\sigma(e^+e^- \to {\rm hadrons},Q)}{\sigma(e^+e^- \to
  \mu^+\mu^-,Q)} \equiv R(Q) = R_{\rm EW}(Q) (1 + \delta_{\rm
  QCD}(Q))$
is sensitive to $\as$ through the $\delta_\text{QCD}$ term
\begin{equation}
  \label{eq:delta-QCD}
  \delta_\text{QCD}(Q) = \frac{\as(Q)}{\pi} + \ldots\,,
\end{equation}
where the series is known up to $\order{\as^4}$~\cite{Baikov:2008jh,Baikov:2012zn}. 
Substituting $\as(\mz) = 0.118$ shows that $\delta_\text{QCD}(\mz)$ is
about a $4\%$ effect, so the roughly per mil measurement
accuracy for $R$~\cite{ALEPH:2005ab}
% cf.
% - R(s=MZ^2) = 1.03904 \pm 0.00087
% - the error on \sigma^0_{had} in table 2, p. 6 of 1407.3792.
%    31.540 \pm 0.037, which is just below per-mil
leads to a $2.5\%$ uncertainty on $\as$.
The actual fits for $\as$, carried out in the context of a global
electroweak fit, are from the GFitter group~\cite{Baak:2014ora},
\begin{subequations}
  \begin{align}
    \label{eq:Gfitter}
    \as(\mz) &= 0.1196 \pm 0.0028_\text{exp} \pm 0.0009_\text{th} =
               0.1196 \pm 0.0030\,.\\
    \intertext{and from the PDG electroweak chapter}
    \label{eq:EW-chapter}
    \as(\mz) &= 0.1203  \pm 0.0028\,.
  \end{align}
\end{subequations}
The two fits differ in the details of which EW input variables are
used and in their error treatment, with the GFitter group
conservatively taking the theoretical uncertainty to be the size of
the last term in the perturbative series.

From the point of view of QCD, the EW fits are arguably the most
robust.
One reason is that the perturbative series is under good control: even
the conservative theory uncertainty from GFitter is below a percent.
The other reason is that non-perturbative corrections are also small,
with $(\Lambda/Q)^4$ (suppressed) or $(\Lambda/Q)^6$ corrections.
Any reasonable estimate of their numerical impact gives contributions
that are much below the experimental uncertainty.
Finally, the extraction is conceptually straightforward, (mostly)
satisfying Guido's criterion of transparency.

A concern that is sometimes raised about the EW extraction of $\as$
(in particular by Guido) is the potential impact of new physics
contributions, such as non-universal vertex corrections, for example
in the $Zb\bar b$ vertex.
For this reason, even if a future collider were to bring improved
experimental accuracy, one might not wish to rely on electroweak fits
alone to extract a precise value for the strong coupling.

%----------------------------------------------------------------------
\section{Tau decays}

The hadronic branching ratio of $\tau$ leptons is an observable that
is sensitive to the strong coupling in a way that is very similar to
the $Z$ width to hadrons, i.e.\ through the slight non-cancellation of
QCD real and virtual graphs.
One practical difference, aside from the much lower momentum scale, is
that in the decay $\tau \to \nu_\tau W^{*}(\to \text{hadrons})$, one
should integrate over all allowed virtualities for the off-shell $W$.
Schematically, following the simplified notation used by Guido, this
gives the following relation for the hadronic branching ratio of the
$\tau$
\begin{equation}
  \label{eq:Rtau}
  R_{\tau} = N \int_{0}^{m_\tau^2} \frac{ds}{m_\tau^2}
  \left(1 - \frac{s}{m_\tau^2}\right)^2 \text{Im}\, \Pi_\tau(s)\,,
\end{equation}
where $N$ is an electroweak normalisation factor and
$\text{Im}\,\Pi_\tau(s)$ is the QCD spectral function.

Experimentally, both $R_{\tau}$ and the detailed spectral function
can be measured, and typically
ALEPH~\cite{Davier:2013sfa,Schael:2005am} and
OPAL~\cite{Ackerstaff:1998yj} data are used, with the former having
somewhat smaller uncertainties.

Theoretically, Eq.~(\ref{eq:Rtau}) involves integrating over squared
hadronic momenta $s$ in a region that has resonance structure and
where perturbative QCD is clearly not applicable.
However, analyticity means that it is possible to rewrite the integral as
\begin{equation}
  \label{eq:Rtau-circle}
  R_{\tau,h} = \frac{N}{2i} \oint_{|s|=m_\tau^2} \frac{ds}{m_\tau^2}
  \left(1 - \frac{s}{m_\tau^2}\right)^2  \Pi_\tau(s)\,.
\end{equation}
The fact that $|s|=m_\tau^2$, together with the $(1-s/m_\tau^2)^2$
factor, ensures that the integral stays away from the most dangerous,
resonance regions.
The theoretical prediction gets evaluated in two main ways: (1) it can
be written as a series in powers of $\as(m_\tau)$, referred to as
fixed-order perturbation theory (FOPT); or (2) one can use the
perturbative expression for $\Pi_\tau(s)$ in the integrand and use the
full renormalisation-group equation for the evolution of $\as(-s)$
around the contour, called contour-improved perturbation theory
(CIPT).\footnote{Yet another scheme is discussed in
  Ref.~\cite{Abbas:2012fi}.}
For some time the choice of FOPT v.\ CIPT was a hotly-debated
one.
Brief and quite approachable explanations of the different points of
view are given in Ref.\cite{Bethke:2011tr}.
Nowadays, most groups tend to quote both and give an average of them
for the final results.

\begin{table}
  %\centering
  \tbl{Extractions of $\as(m_\tau)$ using different choices for both
    the perturbative (PT) evaluation and non-perturbative (NP)
    contributions, based on ALEPH and OPAL data.
    The results are taken from Ref.~\cite{Boito:2016oam}, but the
    central values are indicative also of results from other groups
    when they make similar perturbative and non-perturbative and data
    choices.}
  {\begin{tabular}{@{}ccc@{}}\toprule
    PT choice   & NP choice & $\as^{(n_{\!f}=3)}(m_\tau)$       \\
    \colrule
    FOPT        & DV        & $0.303 \pm 0.009$ \\
    CIPT        & DV        & $0.319 \pm 0.012$ \\
    FOPT        & trunc-OPE & $0.321 \pm 0.009$ \\
    CIPT        & trunc-OPE & $0.339 \pm 0.011$ \\
    \botrule
  \end{tabular}}
  \label{tab:tau-extractions}
\end{table}

Currently the most contentious issue in the literature concerns
non-perturbative corrections.
The two main lines of thought have been exposed recently in 
Refs.~\cite{Pich:2016bdg} (PRS) and \cite{Boito:2016oam}.
Table~\ref{tab:tau-extractions} shows results\footnote{
  Extractions of $\as$ from the $\tau$ data are usually quoted at the
  scale $m_\tau$ with three light flavours.
  To convert $\as^{\smash{(n_{\!f}=3)}}(m_\tau)$ to
  $\as^{\smash{(n_{\!f}=5)}}(m_Z)$, an approximation that is good to
  within about two per mil over the relevant range is
  \begin{equation}
    \label{eq:2}
    \as^{(n_{\!f}=5)}(m_Z) \simeq 0.1180 + 0.125\, [\as^{(n_{\!f}=3)}(m_\tau) - 0.314]\,,
  \end{equation}
  based on 5-loop running~\cite{Baikov:2016tgj,Herzog:2017ohr} and
  four-loop flavour thresholds~\cite{Chetyrkin:2005ia}.
  A given absolute error on $\as(m_\tau)$ goes down by a factor of 8
  when translating it to an error on $\asmz$, in reasonable accord with
  the leading-order expectation that it should scale as $\as^2$.
  This is part of the rationale of extracting $\as$ at such a low scale,
  because the impact of non-negligible non-perturbative effects may be
  still be compensated by the reduction of the error when evolving up to $\mz$.
}
from the Boito et al. paper, with their favoured non-perturbative
choice, which includes ``duality violations'' (DV), i.e.\ an attempt
to allow for differences that may arise between an operator-product
expansion (with quarks, no hadronic resonances) and real data (with
hadrons and corresponding resonances).
It also includes a truncated OPE result, which corresponds to the
approach favoured by PRS, with a more minimal set of non-perturbative
corrections.

PRS argue their approach is justified in part based on the observation
of stability of the extracted $\as$ when replacing the $m_\tau^2$
upper limit of Eq.~(\ref{eq:Rtau}) with an arbitrary limit $s_0$ and
then varying $s_0$.
The PRS final result, averaging the FOPT and CIPT determinations, is
$\alpha_s^{(n_{\!f}=3)}(m_\tau) = 0.328 \pm 0.013$.
However they have considered a range of analyses of non-perturbative
contributions and in some cases found the overall uncertainty can go
up to $0.020$ depending on the precise procedure used (cf.\ their
summary table 11).

\begin{table}
  \tbl{Recent determinations of $\asmz$ from $\tau$ decays that take
    into account the updated ALEPH spectral-function
    data~\cite{Davier:2013sfa} as well as the highest-order
    theoretical predictions~\cite{Baikov:2008jh}.}  {%\footnotesize
    \begin{tabular}{@{}ll@{}}
      \toprule
      Determination & Reference\\
      \colrule
%$0.1202 \pm 0.0019$ &  \arxivonly{Baikov et al.} \cite{Baikov:2008jh}\\[1pt]
$0.1199 \pm 0.0015$ &  \arxivonly{Davier et al.} \cite{Davier:2013sfa}\\[1pt]
$0.1197 \pm 0.0015$ &  \arxivonly{Pich and Rodr\'{\i}guez-S\'anchez} \cite{Pich:2016bdg}\\[4pt]
$0.1175 \pm 0.0018$ &  \arxivonly{Boito et al.} \cite{Boito:2014sta}\\[1pt]
$0.1174\;^{+\;0.0019}_{-\;0.0017}$ &  PDG EW 2016 \cite{Patrignani:2016xqp}\\[1pt]
      \botrule
    \end{tabular}} 
  \label{tab:tau-asmz}
\end{table}

Different results from $\tau$ determinations are summarised in
Table~\ref{tab:tau-asmz}, now showing the values of $\asmz$.
In the case of the Boito et al.\ result, the numbers are taken
from an earlier publication of theirs, however the results are
essentially the same as in their most recent work.

Guido's discussion of $\as$ came before the latest iteration in the
debate about non-perturbative contributions.
Nevertheless he expressed clear concerns, for example about
possible $\order{\Lambda^2/m_\tau^2}$ contributions.
These are absent in the limit of massless quarks, while quark-mass
effects themselves contribute as $m_q^2/m_\tau^2$.
Guido highlighted that if one considers a constituent quark mass,
$m_q \sim 0.3\GeV$, then the resulting $\order{0.3\GeV/m_\tau}^2$ correction to
$R_\tau$ would have a significant impact on the extracted $\as$ value.
Another potential concern is that it is commonplace in $\tau$-based
determinations to vary the renormalisation scale $\mu_R$ by a factor
of $\sqrt{2}$ rather than the more canonical factor of $2$ (though
perhaps this is covered by the CIPT/FOPT difference). 

My inclination is to share Guido's caution about determinations of
$\as$ from $\tau$ decays, especially in view of the ongoing debates on
the subject.
One question is whether to view $\tau$ decays as a source of precise
information about $\as$ or rather a unique window into physics close
to the edge of the perturbative regime.

%----------------------------------------------------------------------
\section{PDF determinations}

PDF fits are sensitive to the strong coupling in various ways.
One way is through the $Q^2$ dependence of Deep Inelastic Scattering
(DIS) structure functions, itself driven by the
DGLAP~\cite{Gribov:1972ri,Altarelli:1977zs,Dokshitzer:1977sg}
evolution of the underlying PDFs, which is proportional to the
coupling,
\begin{equation}
  \label{eq:3}
  \frac{d q(x,Q^2)}{d \ln Q^2} = \as(Q^2) \int_x^1 \frac{dz}{z}
  (P_{qq}(z) q(x/z, Q^2) + P_{qg}(z) g(x/z,Q^2))
  + \order{\as^2},
\end{equation}
where $P_{ij}$ are splitting functions and $q$ and $g$ are quark and
gluon distributions.
Another source of sensitivity is that some cross sections are
proportional to $\as$ or $\as^2$, e.g.\ jet cross sections, which
overlaps with the question of collider determinations below.

\begin{table}
  \tbl{Determinations of the strong coupling in the context of PDF
    fits.
    The first uncertainty is the statistical uncertainty of the fit as
    defined by the fit authors (scaled from $90\%$ to $68\%$ confidence
    level in the case of CT14), the second is a theory uncertainty
    estimated as half the difference between the NLO and NNLO fit
    results.
    In the case of ABMP16, which carried out only a NNLO fit, the
    NLO$-$NNLO difference is taken from the earlier, conceptually
    similar, ABM12 fit~\cite{Alekhin:2012ig}.
    The JR result corresponds to their ``standard'' fit variant, which
    has minimal assumptions about the structure of parton distributions
    at low $Q^2$.  } {%
    \begin{tabular}{@{}lll@{}}
      \toprule
      Determination & PDF fit & Reference\\
      \colrule
$0.1141\;^{+\;0.0022}_{-\;0.0020}\; \pm 0.0003$ &  BBG06 non-singlet & \cite{Blumlein:2006be}\\[1pt]
$0.1173 \pm 0.0007 \pm 0.0009$ &  NNPDF21 & \cite{Ball:2011us}\\[1pt]
$0.1162 \pm 0.0006 \pm 0.0014$ &  JR14 & \cite{Jimenez-Delgado:2014twa}\\[1pt]
$0.1147 \pm 0.0008 \pm 0.0023$ &  ABMP16 & \cite{Alekhin:2017kpj}\\[1pt]
$0.1150\;^{+\;0.0036}_{-\;0.0024}\; \pm 0.0010$ &  CT14 & \cite{Dulat:2015mca}\\[1pt]
$0.1172 \pm 0.0013 \pm 0.0014$ &  MMHT2014 & \cite{Harland-Lang:2015nxa}\\
      \botrule
    \end{tabular}} 
  \label{tab:PDFs}
\end{table}

A summary of extractions of $\as$ carried out in the context of PDF
fits is given in Table~\ref{tab:PDFs}, taking only the most recent
published result from any given fitting group and/or approach.
One element to note is that in contrast to almost all other classes of
strong coupling determination, PDF fits don't usually quote a theory
uncertainty.
Partly this is because it is not straightforward to extend the
standard method for theory uncertainty estimation, scale variation, to
PDF fits: many different processes come into play and in each one
scale variations effectively play a different role.
One then ends up with hard-to-answer questions of whether scale
variations should be correlated across processes and even across
different regions of $x$ and $Q^2$.
Neglecting theory uncertainties in the PDF fit means that if two
processes or kinematic regions have similar statistical constraining
power but one has much larger theory uncertainties, the region with
larger theory uncertainties will get more weight than is
appropriate.\footnote{As far as I'm aware, the extent to which this
  situation occurs in practice hasn't been studied in detail.}

A poor-man's approach to estimating theory uncertainties is to take
half the difference between fits at NNLO and NLO, as adopted by the
NNPDF collaboration.
This is the basis of the theory uncertainties shown in
Table~\ref{tab:PDFs}.\footnote{The BBG06 result is based on N$^3$LO
  coefficient functions and in that case the table shows half the
  difference between N$^3$LO and NNLO.
  At the time of its publication only NNLO splitting functions were
  available, however the authors argue that the uncertainty from the
  N$^3$LO splitting function was small relative to other uncertainties
  and that this statement is further supported by the recent
  calculation of the exact N$^3$LO non-singlet splitting
  functions~\cite{Moch:2017uml}.
  I am grateful to Johannes Bl\"umlein for correspondence on this
  point. 
}
Formally it's a conservative approach (the theory uncertainty
estimated in this way is of the same order as the NNLO corrections),
though in practice it might underestimate the error if NNLO and NLO
results just happen, numerically, to be close.

Taking into account both the experimental and the estimated theory
uncertainties, the different PDF determinations are largely consistent
with each other.
This has not, however, prevented heated debate between groups.
For example if ones leaves aside the theory uncertainty estimate and
assumes the experimental errors to be uncorrelated, the relatively
recent MMHT2014 and ABMP16 results are $1.7\sigma$ apart.
In reality the experimental errors should be correlated, since much of
the underlying data is the same.
So the disagreement must be ascribed to systematic differences between
the fit procedures.
These include: the treatment of heavy flavour, whether a fixed-flavour
number scheme as in ABMP16 or a general-mass variable flavour number
scheme (GM-VFNS) as in MMHT2014 (as well as CT14 and NNPDF21);
the treatment of higher twist effects, with ABMP16 explicitly
including higher-twist terms within their cross-section calculations,
while many other groups don't;
the inclusion of collider jet data only at NLO, which is
inconsistent with the rest of the fit being NNLO, an issue whose
importance was debated given the non-negligible experimental
uncertainties of the datasets.
Each group argues that its results are robust~(see e.g.\
Refs.~\cite{Alekhin:2013nda,Thorne:2014toa}). 

Guido expressed a preference for the results based on global fits,
i.e.\ using the largest available data sets, which today corresponds
to the CT, MMHT and NNPDF results.
Those also represent my first choice, mainly because of their use of a
GM-VFNS which is relevant for the moderate and high $Q^2$ DIS data.
They also represent the widely adopted choice of the LHC community
for the PDFs themselves, notably through the PDF4LHC15 combined PDF
set~\cite{Butterworth:2015oua} (which does not involve an $\as$
fit).

Looking to the future, there are prospects for significant theoretical
improvements in such fits, for example from the full NNLO jet cross
sections~\cite{Currie:2016bfm} already recently included into a first
PDF fit~\cite{Harland-Lang:2017ytb} or the inclusion of $Z$ $p_t$
distribution data~\cite{Boughezal:2017nla}.\footnote{Beware, however,
  of the effect of $\Lambda/Q$ power corrections for these datasets.
  In particular the pattern of soft gluon emission from a $Z$+jet event
  is azimuthally asymmetric: there is more radiation away from the $Z$
  than in the same direction as the $Z$.
  As a result one may expect non-perturbative modifications of the
  pattern of emission to also be azimuthally asymmetric, resulting in
  a net average shift of the $Z$ $p_t$ by an amount of order
  $\Lambda$.
  A $0.5\GeV$ shift would translate to a $1.5\%$ change in the $Z$ $p_t$
  distribution around $p_t = 100\GeV$, which is significant compared
  to the sub-percent accuracies of some measurements~\cite{Aad:2015auj}.
  This type of effect cannot be straightforwardly estimated by turning
  hadronisation and multiple-parton-interaction effects on and off in
  Monte Carlo simulations.  }
Other advances that could be of benefit to all fits include small-$x$
resummation (as used in Ref.~\cite{Ball:2017otu}) and recent progress
towards N$^3$LO splitting functions~\cite{Moch:2017uml}.

However, I do have concerns about potential fundamental limits in PDF
fits as they are carried out currently.
In the case of $\tau$ decay we saw there is
considerable debate about non-perturbative corrections, for a
kinematic region $s = m_\tau^2 = 3.16\GeV^2$.
DIS fits extend to a comparably low $Q^2$, but power corrections
rather than being $\Lambda^p/Q^p$ with $p=4$ or $6$, have $p=2$, i.e.\
they are potentially larger.
Additionally, charm production is a relevant contribution to the $F_2$
structure function and it is not clear how reliably it can be
predicted near threshold (even within a minimal ``fitted'' charm
framework), given the hundreds of MeV difference between a charm quark
mass and charm-meson masses and the significant mass-dependence of the
cross sections.

How could these problems be addressed?
Some PDF fits already explore the use of higher $Q^2$ cutoffs on the
data being used and I think this is an avenue that deserves to be
pursued further.
For example, one might argue that a relative precision $\epsilon$ on
PDFs, and the associated $\as$, is to be trusted only insofar as the
DIS data being used satisfies
$Q^2 > Q_\text{min}^2 \gtrsim \Lambda^2 / \epsilon$.
The choice of $\Lambda^2$ would need to be debated, but could be of
the order of $0.5\GeV^2$.

%----------------------------------------------------------------------
\section{Collider determinations}

By ``collider determinations'' I mean determinations of $\as$ based on
cross sections measured at hadron--hadron and hadron--lepton colliders
that are used to constrain the strong coupling independently of a PDF
fit.

With the recent advances in NNLO calculations (see a recent
review~\cite{Dittmaier:2017vus}), a significant number of new
processes is becoming available for collider-based NNLO
strong-coupling determinations.

In practice two processes have been used so far, $t\bar t$
production~\cite{Chatrchyan:2013haa,Klijnsma:2017eqp} together with the
calculation of Ref.~\cite{Czakon:2013goa}, giving
\begin{equation}
  \label{eq:ttbar}
  \as(\mz) = 0.1177
  \pm 0.0010 \,(\text{exp.})\,
  ^{+0.0020}_{-0.0024}\, (\text{PDF})\,
  ^{+0.0021}_{-0.0021}\, (\text{scale})\,
  = 0.1177 
  ^{+0.0034}_{-0.0036}\,, 
\end{equation}
which combines a number of top-production cross section measurements
from ATLAS, CMS and the Tevatron;
and jet production in DIS~\cite{Andreev:2017vxu} by the H1
collaboration, together with the calculation
Ref.~\cite{Currie:2016ytq},
\begin{align}
  \label{eq:DIS-jets}\nonumber
  \as(\mz) &= 0.1157
  \pm 0.0020 (\text{exp})
  \pm 0.0006 (\text{had})
  \pm 0.0005 (\text{PDFs})
  \pm 0.0027 (\text{scale})\,,\\
  &= 0.1157 \pm 0.0034\,,
\end{align}
where the PDF uncertainty includes several sources: the actual PDF
uncertainty, the dependence of the PDF on $\as$ and differences
between PDF sets.
These uncertainties are all small, perhaps because the jet production
kinematic region that was used is dominated by quark-induced
processes.

One potential question about individual collider determinations is why one would
go down this route at all: isn't it better simply to include the
collider data in a global PDF fit and extract $\as$ that way?
One answer to this is that it is much simpler to properly account for
scale and other theoretical uncertainties when considering a single
observable than when considering many different observables and kinematic
ranges.
What's more, one sees that the theory uncertainties tend to be as
large as any other uncertainty: neglecting them, as is common in
global PDF fits, is clearly not justified.%
\footnote{The H1 paper also shows the result of extracting $\as$
  within a PDF fit that incorporates the H1 structure function and jet
  data and find $\as(\mz)= 0.1142 \pm 0.0028$.
  Interestingly this fit includes scale variations both for the
  structure functions and the jet cross section and it is the scale
  variation uncertainty ($\pm0.0026$) that dominates the final $\as$
  uncertainty.
  The fit also restricts its attention to $Q^2 > 10\GeV^2$, which is
  further from the dangerous non-perturbative region than
  the $Q_\text{min}^2$ value used in many global fits.
  %
  % It uses a ZM-VFNS, but maybe this is OK for these $Q^2$ values?
}

Overall, even if I've been involved in them myself, I am inclined to
approach collider $\as$ fits with some caution.
Ultimately, insofar as they rely on knowledge of PDFs, they inherit
the same drawbacks as PDF fits, notably the potential sensitivity
to low $Q^2$ non-perturbative effects.
Only if one can devise sets of collider observables where the
sensitivity to PDFs is mostly eliminated can one evade this problem.
To some extent this sensitivity to PDFs is eliminated in the H1 study.
However the use of a cut of
$\tilde \mu^2 \simeq Q^2 + p_{t,jet}^2 > (28 \GeV)^2$ means that some
fraction of the jets will have a transverse momentum that is
sufficiently low that ($\Lambda/p_t$) hadronisation effects could be a
concern beyond the quoted hadronisation uncertainty, as was the case
for event-shape fits.

%----------------------------------------------------------------------
\section{Lattice QCD}

Whereas most $\as$ determinations involve comparing a perturbative
calculation directly with an experimental observable, lattice QCD
approaches use a somewhat different methodology:
firstly, lattice parameters are tuned so as to reproduce suitably
chosen low energy hadronic data (e.g.\ pion and kaon decay constants);
then the same lattice simulation is used to calculate some observable
at a perturbative scale that is also amenable to calculation in
perturbation theory;
finally $\as$ is determined by requiring agreement between the lattice
and perturbative predictions for that observable.

The main recent lattice results for $\as$ (using at least $2+1$
flavours) are summarised in Table~\ref{tab:lattice}.
They differ from each other both in the type of lattice simulations
used (e.g.\ the treatment of light quarks) and in the choice of
observable used to match with perturbation theory.

\begin{table}
  \tbl{Selected determinations of the strong coupling within lattice
    QCD.}  {%\footnotesize
    \begin{tabular}{@{}lll@{}}
      \toprule
      Determination & Approach & Reference\\
      \colrule
$0.1184 \pm 0.0006$ &  HPQCD Wilson loops & \cite{McNeile:2010ji}\\[1pt]
$0.1192 \pm 0.0011$ &  Maltman-HPQCD Wilson loop & \cite{Maltman:2008bx}\\[4pt]
$0.1182 \pm 0.00074$ &  HPQCD heavy quark current correlator & \cite{Chakraborty:2014aca}\\[1pt]
$0.1177 \pm 0.0026$ &  JLQCD charmonium correlators & \cite{Nakayama:2016atf}\\[1pt]
$0.1162 \pm 0.00084$ &  MP charmonium correlators & \cite{Maezawa:2016vgv}\\[4pt]
$0.1196 \pm 0.00108$ &  ETM ghost-gluon coupling & \cite{Blossier:2013ioa}\\[1pt]
$0.1166\;^{+\;0.0012}_{-\;0.0008}$ &  QCD static energy & \cite{Bazavov:2014soa}\\[4pt]
$0.1205\;^{+\;0.00094}_{-\;0.00197}$ &  PACS-CS step scaling & \cite{Aoki:2009tf}\\[1pt]
$0.1185 \pm 0.00084$ &  ALPHA step scaling & \cite{Bruno:2017gxd}\\      \botrule
    \end{tabular}} 
  \label{tab:lattice}
\end{table}

The first high-precision $\as$ results with reasonable dynamical quark
masses were those from the HPQCD
collaboration~\cite{Davies:2008sw,McNeile:2010ji,Chakraborty:2014aca},
with precisions approaching $0.5\%$.
The solidity of this precision claim has been widely discussed in the
literature.
Guido's comment was
\begin{quotation}
  \noindent With all due respect to lattice people I think this small error is
  totally [i]mplausible.
\end{quotation}
The lattice part of the calculation uses the staggered fermion
approach for the light quarks, a method that is the subject of
misgivings by part of the lattice community.
The perturbative correspondence is made using Wilson loops and
heavy-quark correlators.
Concerns have been raised as to whether perturbation theory is
sufficiently precise for these observables at the scale where $\as$ is
extracted.
The HPQCD authors argue that they have been conservative in
their estimate of perturbative matching systematics.
However the method involves fitting higher-order coefficients in the
perturbative series, an approach that is not widely used in other
$\as$ determinations.
With a subset of the same lattice data, but different assumptions in
using it, Ref.~\cite{Maltman:2008bx} found a slightly different result
with almost double the uncertainty.
The FLAG working group~\cite{Aoki:2013ldr}, representing part of the
community, argued that a larger error uncertainty should be assigned
to the HPQCD results, notably associated with the matching with
perturbation theory and its final estimate for $\asmz$ was
$0.1184 \pm 0.0012$.

The question of the reliability of perturbation theory applies to
essentially all of the determinations shown in
Table~\ref{tab:lattice}, and in many cases there is also a concern
about the potential impact of discretisation errors.
This is because of the fundamental computational limitation, within
any single lattice calculation, on the ratio of the smallest length
scale to the longest length scale.
The former needs to be as small as possible for good perturbative
matching and small discretisation errors at the perturbative scale,
while the latter needs to be large to minimise finite-volume effects in
low-energy observables.
The issue of the perturbative scale being insufficiently high leads to a
situation where different groups, using broadly similar methods,
obtain substantially different error estimates.
This is the case for the JLQCD v. MP charmonium correlator results:
the JLQCD result uses scale variation to estimate the perturbative
uncertainty and this dominates the $2.2\%$ uncertainty.
%
%% For MP: see discussion in whole paragraph after Eq.(29)
In contrast the MP result varies an unknown higher order term in the
perturbative series, within some estimated reasonable range, and finds
a negligible contribution to the overall $0.7\%$ uncertainty, which
is instead dominated by continuum extrapolation and statistical
components.

A notable advance in this respect has come recently from the ALPHA
collaboration~\cite{Bruno:2017gxd}.
Here the matching is performed at a scale of $70\GeV$, where $\as$ is
sufficiently small that systematic errors from neglected higher order
terms are convincingly subdominant~\cite{Brida:2016flw}.
The authors obtain lattice results at this high scale with the help of
the step scaling method~\cite{Luscher:1991wu}, i.e.\ by using a series
of lattice simulations with progressively smaller sizes, i.e.\ higher
momentum scales, non-perturbatively matching each lattice to the
previous one.\footnote{An earlier step-scaling analysis was carried
  out by the PACS-CS collaboration~\cite{Aoki:2009tf} and while it was
  relatively far from the chiral limit, its results are consistent
  with those from the ALPHA collaboration.}
Their final error of $0.7\%$ is dominantly statistical.

\begin{figure}[tp]
  \centering
  \includegraphics[width=\textwidth]{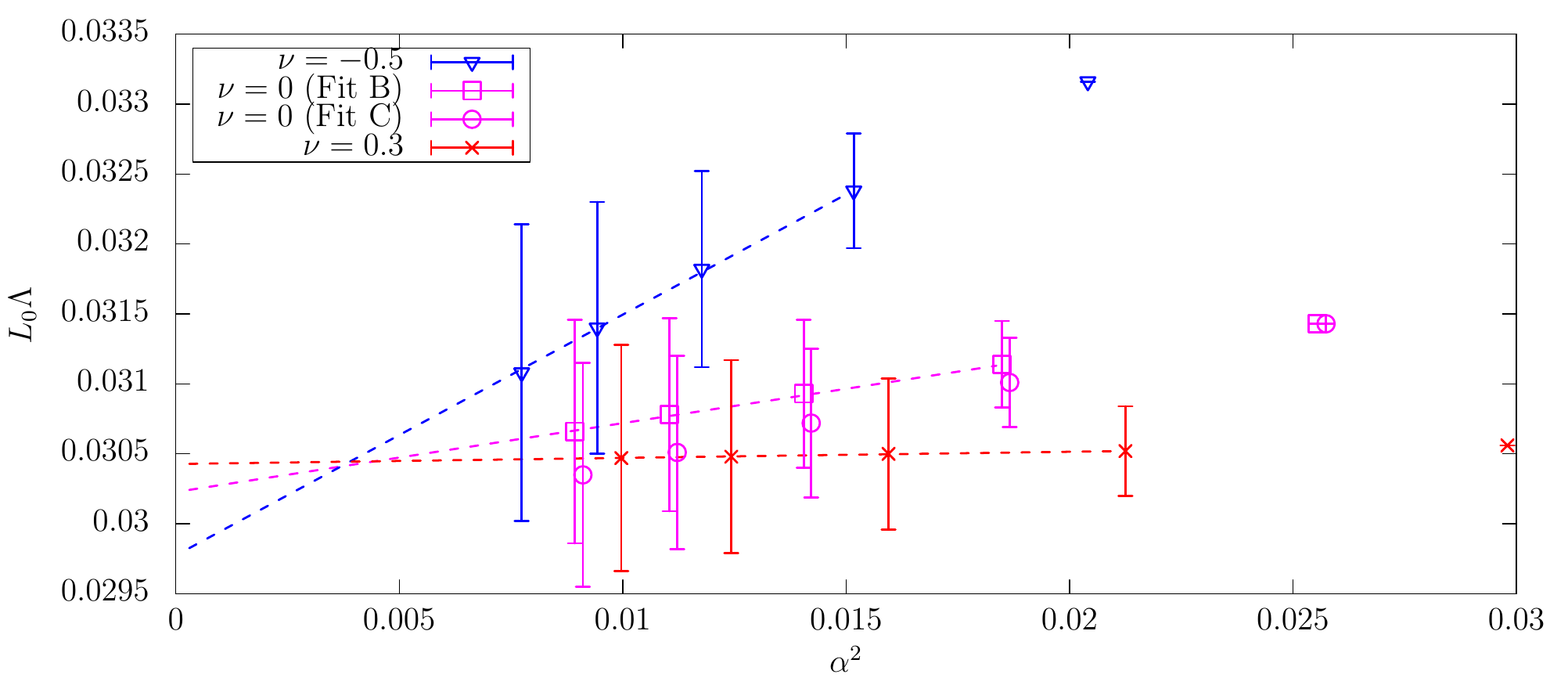}
  \caption{
    Adaptation of Fig.~2 of Ref.~\cite{Brida:2016flw}, kindly
    provided by Alberto Ramos.
    The figure
    shows the extraction of the QCD $\Lambda$ parameter, in
    units of $L_0 \simeq 4.22 \GeV$, in the Schr\"odinger Functional scheme.
    Different lattice sizes (or correspondingly scales $\mu_\text{PT}$
    in the step scaling approach) each lead to one
    $\as(\mu_\text{PT})$, the value of which is shown squared on the
    $x$ axis (labelled $\alpha^2$). 
    For each $\as(\mu_\text{PT})$ result, the authors deduce a value
    for the QCD $\Lambda$ parameter ($y$ axis).
    The parameter $\nu$ in the underlying Schr\"odinger Functional
    approach can be thought of as a renormalisation scheme choice
    ($\Lambda$ is always converted back to a unique scheme).
    The two $\nu=0$ sets of results (labelled Fit B, C) correspond to different
    correction schemes for lattice discretisation artefacts.
  }
  \label{fig:lambda-from-alpha}
\end{figure}

A demonstration of the perturbative robustness of the ALPHA
determination is given in Fig.~\ref{fig:lambda-from-alpha}, which
shows the extraction of the QCD scale $\Lambda$ (defined in terms of $\as(\mu)$ in Eqs.~(1,2) of
Ref.~\cite{Brida:2016flw}\footnote{In most circumstances, the PDG
  recommends against referring to values for $\Lambda$.
  This is because in the collider community $\Lambda$ is often used as
  parameter in closed-form formulas for $\as$ that do not exactly
  satisfy the renormalisation group equation to some truncated order
  in the beta-function.
  A same value for $\Lambda$ can also lead to different $\as$ values
  depending on the specific closed-form formula used.
  In contrast the definition of $\Lambda$ that is exploited in
  Ref.~\cite{Brida:2016flw} is defined through an implicit equation for
  $\as$ that does exactly satisfy the renormalisation group
  equation.}), %
based on a number of differently sized lattices, covering a factor of
16 in scales.
For each lattice size, they determine a value of $\as$ at a scale
$\mu_\text{PT}$ associated with the lattice size.
Then using purely perturbative running they can convert it to a value
for $\Lambda$.
This is done for a variety of schemes, corresponding to the choices of
$\nu$ in the figure (the $\Lambda$ value shown is defined in a scheme
invariant way).
With a $\beta$-function known to all orders and in the absence of
power corrections, the extracted $\Lambda$ would be independent of
$\mu_\text{PT}$ or correspondingly of the $\as^2(\mu_\text{PT})$ value
shown on the $x$ axis.
In practice, missing higher order terms for the $\beta$ function
in the Schr\"odinger Functional scheme should correspond to a
residual offset for the extracted $\Lambda$ value that scales as
$\alpha_s^2(\mu_\text{PT}^2)$, with a $\nu$-dependent coefficient.
This is observed in the lattice data over the last three iterations of
step scaling, insofar as the four leftmost points (covering a factor
of $8$ in $\mu_\text{PT}$) are consistent with a linear dependence of
the extracted $\Lambda$ parameter on $\as^2$.
An extrapolation of $\Lambda$ to zero coupling is always consistent
with the result at the lowest available value of $\as$, within its
statistical errors.
Furthermore, at that lowest value of $\as$, the different schemes are
all in agreement.
This is a powerful cross-check of the stability of the perturbative
side of the extraction over a broad range of scales.
The final uncertainty of $2.6\%$ on $L_0 \Lambda$ that is quoted by the
authors corresponds to about $0.54\%$ uncertainty on $\asmz$.
The conversion of these results to an $\overline{\text{MS}}$ $\Lambda$
(or coupling) in physical units, in particular the determination of
$L_0$ in GeV, corresponds to the work of Refs.~\cite{DallaBrida:2016kgh,Bruno:2017gxd}.

One issue raised by the ALPHA collaboration is that their results (as
most others) are for $2+1$ flavours, and so require perturbative
matching at the charm mass in order to be interpreted as a 5-flavour
coupling at $\mz$.
However, the 2015 HPQCD result with $2+1+1$
flavours~\cite{Chakraborty:2014aca} is very close to the earlier
result with $2+1$ flavours and a perturbative charm
threshold~\cite{McNeile:2010ji}.
This suggests that non-perturbative charm-threshold effects are small.
A further potential concern is that QED and isospin breaking effects
are not accounted for in the lattice simulations of the low-energy
hadronic quantities, however these are believed to be subdominant
compared to the current uncertainties.
Nevertheless, such points may need to be addressed in future
higher-accuracy determinations of the strong coupling.

%----------------------------------------------------------------------
\section{Concluding remarks}

\begin{figure}[tp]
  \centering
  \includegraphics[width=0.69\textwidth]{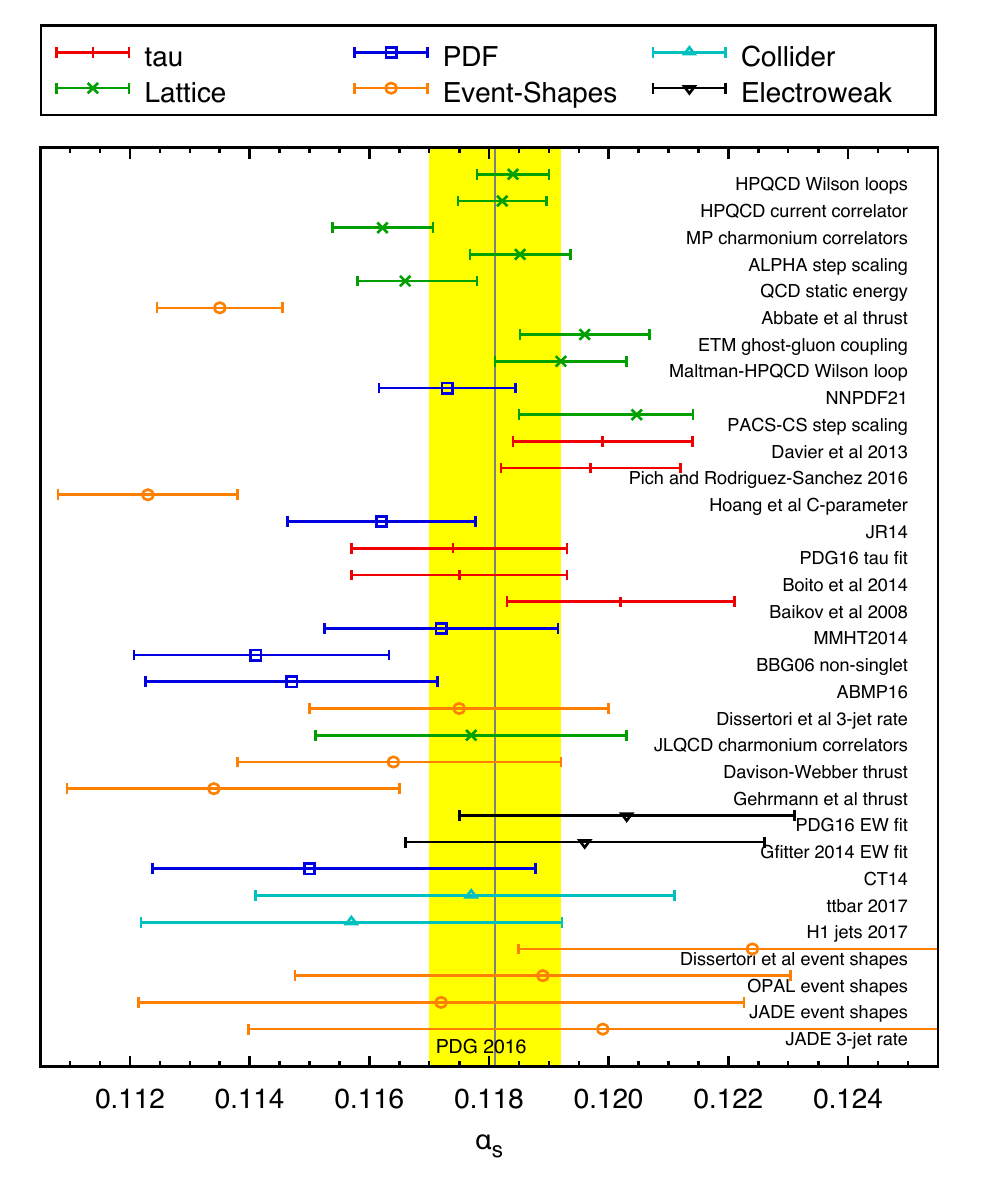}%
  \caption{The $\as$ determinations discussed in this review, in order
    of increasing quoted uncertainty.}
  \label{fig:summary}
\end{figure}

The many different (NNLO and better) determinations are summarised in
Fig.~\ref{fig:summary}.
A wide range of methods can deliver results for $\as$ with an accuracy
of a few percent.
However such an accuracy is not sufficient if one is to fully exploit
the results that are coming and still to come from the LHC.
The question then for the field is which, if any, of the percent-level
determinations to trust given that some of them are mutually
incompatible.
A few years ago, Guido's answer to that question was, essentially,
none!

The key issues that recur are the estimate of uncertainties from
missing higher orders and the problem of non-perturbative corrections.
Many of the determinations based on $\tau$ decays, event shapes, PDF
fits and lattice methods are either directly determining $\as$ at a
scale in the range of $1{-}3\GeV$ or are in some other way sensitive
to the physics occurring on those scales.
Most of them give detailed reasoning as to why they believe they are
able to control the problems that might arise from proximity to the
non-perturbative region and the poor convergence of the series when
$\as$ is not so small.
However the accessible range of scales over which one can rigorously
test these statements tends to be limited, leaving the door open to
sometimes heated debate about the degree of control over systematic
uncertainties.

In this respect the recent ALPHA lattice result is potentially a
breakthrough.
Together with EW precision fits, it is the only approach where the
connection with perturbation theory is made at a genuinely high scale,
with almost no assumptions needed about low-scale physics.
Its precision of $0.7\%$ is almost three times better than the result
from EW fits, thanks in part to the fact that it is not limited by LEP
statistics.
Furthermore the method can provide an extraction over a wide range of
momentum scales and the pattern of results obtained in that way gives
strong support to the authors' arguments that the perturbative
extraction is robust.

What would be needed to consolidate the ALPHA advance?
Any determination of $\as$ that reaches percent-level accuracy tends
to be a complicated endeavour, with aspects that can be adequately
judged only by experts in that sub-field.
That goes against the criterion of transparency called for by Guido.
In such situations, sometimes the best way of judging a result is to
attempt to reproduce it, ideally making complementary choices where
relevant.
A second, independent, up-to-date lattice step-scaling determination
would thus be important to help establish this approach as the
reference method for $\as$ determinations.

In the meantime, what global average should one use?
For the 2017 PDG update we decided to maintain the 2016 average,
$\asmz = 0.1181 \pm 0.0011$ in order to allow more time to collect
community input on the latest $\as$ determinations.
This world average remains consistent with, if somewhat more
conservative than, a simple weighted average of the two determinations
that appear today to be the cleanest, namely the ALPHA and EW
determinations, which gives $\asmz = 0.1186 \pm 0.0008$.

%----------------------------------------------------------------------
\section*{Acknowledgements}

Aside from those with Guido, I have benefited from discussions on
$\as$ with many colleagues and friends over the years.
In particular I would like to thank Siggi Bethke and G\"unther
Dissertori for the past decade of collaboration on the PDG QCD review
chapter, including discussions about all aspects of $\as$
determinations (and together with Thomas Klijnsma also our work on
$\as$ from top-quark cross sections), as well as Jens Erler for
regular discussions about aspects that overlap with the Electroweak
review;
Martin L\"uscher, Luigi del Debbio and Agostino Patella, for numerous
discussions about lattice approaches and Alberto Ramos for extensive
exchanges on the ALPHA determination and for supplying
Fig.~\ref{fig:lambda-from-alpha}; 
Martin Beneke and Toni Pich for their insights on $\tau$
determinations (if my conclusions differ from theirs it may simply be
because I still have more to learn), as well as Andreas H\"ocker and
Zhiqing Zhang for clarifications;
Johannes Bl\"umlein, Stefano Forte, Sven Moch, Juan Rojo and Robert
Thorne for discussions about PDF fits;
Andrea Banfi, Stefan Kluth, Gionata Luisoni, Pier Monni
and Giulia Zanderighi for their insights about $e^+e^-$ jet rates and
event shapes and their input for Fig.~\ref{fig:aleph-jet-rate} as well
as G\"unther Dissertori for permission to reproduce the top-right
panel of Fig.~\ref{fig:aleph-jet-rate};
and, finally, Zolt\'an Tr\'ocs\'anyi for discussions concerning the
interplay of the strong coupling and vacuum stability.
I would also like to thank Siggi Bethke, Michelangelo Mangano, Pier
Monni, Alberto Ramos and Giulia Zanderighi for helpful comments on the
manuscript.

%%% Local Variables:
%%% TeX-master: "guido-book-holder.tex"
%%% End:

%  LocalWords:  Altarelli logarithmically electroweak virtualities Eq
%  LocalWords:  datasets discretisation correlator charmonium Siggi
%  LocalWords:  Bethke unther PDG ggHiggs ABMP ALEPH Pythia DW Monash
%  LocalWords:  GeV integrable MChad NLL anlhad SCET antiquark hadr
%  LocalWords:  Guido's Gionata Luisoni Monni EW SLC GFitter FOPT DV
%  LocalWords:  CIPT Boito PRS et al ij BBG NNPDF MMHT VFNS MeV kaon
%  LocalWords:  HPQCD Maltman JLQCD ETM PACS chiral Eqs Schr odinger
%  LocalWords:  isospin Dissertori Jens Erler Debbio Agostino Beneke
%  LocalWords:  Pich Andreas Zhiqing Zhang Stefano Moch Rojo Thorne
%  LocalWords:  Banfi Kluth Giulia Zanderighi th lll Zb ds Im Mangano
%  LocalWords:  analyticity Klijnsma

\bibliographystyle{JHEP}
\bibliography{gavin}

\end{document}